\documentclass[aps,twocolumn,showpacs,superscriptaddress,
floats,prb]{revtex4}

\usepackage[english]{babel}
\usepackage[latin1]{inputenc}
\usepackage{amsmath, amssymb, amsfonts}
\usepackage{graphicx,epsfig,dsfont}
\usepackage{float}

\usepackage[squaren]{SIunits}
\usepackage{color}
\usepackage{bbm}

\newcommand{\Z}{{\mathbb{Z}}}
\newcommand{\HH}{{\mathcal H}}
\newcommand{\id}{{\mathbbm{1}}}
\newcommand{\T}{{\mathcal T}}
\newcommand{\Q}{{Q}}

\begin{document}

 \title{Unwinding of a one-dimensional topological superconductor}

\author{Achim Rosch}
\affiliation{Institute for Theoretical Physics, University of Cologne, 
D-50937 Cologne, Germany
}

\begin{abstract}
  We show that a topological superconductor made of four chains of
  superconducting {\em spinless} fermions characterized by four
  Majorana edge states can adiabatically be deformed into a trivial
  band insulator. To unwind this time-reversal invariant topological
  superconductor, interactions to {\em spinful} fermions are switched on
  along an adiabatic path. Thereby, we couple modes which belong to
  two different representations of the time-reversal symmetry operator
  $\T$ with ${\T}^2=1$ and ${\T}^2=-1$, respectively. This observation
  can be understood by investigating how the relevant
  symmetries act on the entanglement spectrum giving 
rise to {\em four} instead of {\em eight}
  different topological phases with Majorana edge modes. We
  also show that a simple level crossing of doubly and singly
  degenerate states occurs in the entanglement spectrum upon deforming
  the quantum state.
\end{abstract}

\date{\today}

\maketitle
\section{Introduction}
The concepts of symmetry and topology allow to classify states of matter.
A famous example is the quantum Hall effect, where the quantization of the Hall conductance is directly connected to the existence of topological protected edge states which are robust against disorder and interactions. Only more recently, it was realized \cite{Kane2005, Bernevig2006,Fu2007, Moore2007, Roy2009} that similar topological states with protected edge modes also exist in time-reversal invariant systems in three dimensions, which led to a considerable excitement in the field \cite{rev1,rev2} and intensive theoretical and experimental \cite{Koenig2007,Hasan2009, Bruene2011} investigation of such systems. A major achievement was the systematic topological classification for non-interacting systems \cite{schnyder08,zhang08,kitaev09} under the assumption that only a certain set of discrete symmetries (time-reversal invariance and various particle-hole symmetries relevant, e.g., for superconductors \cite{altland}) are present.

An obvious question is to what extent this classification can also be used to classify interacting systems. In some cases such as the quantum Hall effect or the three-dimensional topological insulator this question can be answered as it is possible to identify measurable bulk properties (the Hall conductance or the $\theta$ term \cite{Qi2008}, respectively) which are quantized both in the presence and absence of interactions. For other cases the situation is less clear. 

Recently,  Fidkowski and Kitaev in an insightful paper \cite{fidkowski}  identified a case where a different topological classification was obtained for interacting and non-interacting systems. In one dimension, they considered chains of time-reversal invariant superconducting spinless fermions (class BDI \cite{altland,schnyder08,kitaev09}). In the absence of interactions, such chains are classified by an integer $n \in \Z$ which physically describes the presence of $n$ Majorana states $\alpha_i$, $i=1,...,n$, at a given edge which all have the same signature under time-reversal $\T$. Therefore the hermitian quadratic coupling terms, $i \alpha_j \alpha_{j'}$, are forbidden as they are not time-reversal invariant. In contrast, interaction terms of the form $\alpha_1 \alpha_2 \alpha_3 \alpha_4+H.c.$ are allowed. For $n=8$ they can lead to a unique ground state at the edge. This fact allowed Fidkowski and Kitaev to construct explicitly an adiabatic path to connect the $n=8$ state to the trivial insulator. Thereby they showed that in the presence of interactions the classification in terms of $\Z$ has to be replaced by $\Z_8$. Later insightful papers by Turner, Pollmann and Berg \cite{turner11} and also Fidkowski and Kitaev \cite{fidkowski2}   showed how these results can be understood by investigating how the symmetries of the system affect  the entanglement spectrum \cite{turner11,fidkowski2} (see also \cite{wen11,cirac11} for the complete classification of 1d bosonic systems using projective symmetries) obtained by calculating the eigenvalues of the reduced density matrix after tracing out part of the system.

\begin{figure}[t]
\begin{center}
\includegraphics[width=\linewidth,clip]{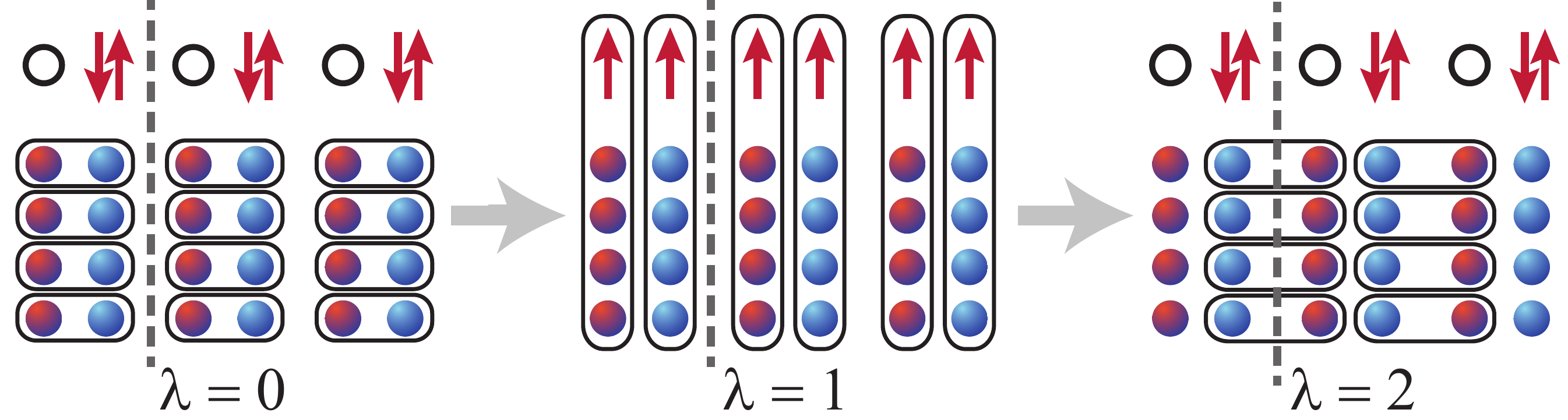}
\end{center}
\caption{(color online) Schematic sketch of the adiabatic transformation from a trivial insulator (left, $\lambda=0$) to a superconducting state characterized by four Majorana edge states on each side of the open chain (right, $\lambda=2$). During this transformation the coupling to spinful fermions (upper line) is switched on. At $\lambda=1$ a pseudospin made from 4 Majoranas forms a spin-singlet with the spin of an electron.  \label{figtrans}}
\end{figure}
The $\Z_8$ topological classification of interacting spinless and
superconducting fermions in Refs.~[\onlinecite{fidkowski2,turner11}]
is based on two symmetries, conservation of particle number parity,
$Q=(-1)^N$, and time-reversal symmetry, $\T$, and three important
equations, $\T^2=1$, $Q^2=1$ and $[Q,\T]=0$. It is therefore not
surprising, that if one of these conditions, $\T^2=1$, is relaxed, the
classification scheme has to change. Indeed we will show that when we
allow a coupling of the spinless fermions to spinful fermions (with
$\T^2=\pm 1$ for even/odd numbers of spinful fermions)  it is possible to deform a state with 4 Majorana edge modes into the trivial insulator without closing the band gap, see Fig.~\ref{figtrans}. 

A related result has been obtained a few years ago by us\cite{anfuso}
when considering the topological stability of spin-1 Haldane chains
and of various gapped phases of spin-1/2 ladder models. At the edge of
a gapped spin-1 Haldane chain, localized spin-1/2 states emerge. Those
are topologically stable, if one considers pure spin models with an integer total spin on each rung of the ladder. If,
however, one remembers that in experimental system spin-1 degrees
of freedom arise typically from the spin of two ferromagnetically coupled
electrons, one should ask the question, whether the topological
stability persist, when one considers the larger Hilbert space
including the fermionic degrees of freedom. Here we showed by an
explicit construction\cite{anfuso} that one can adiabatically
transform the wave function of a spin-1 Haldane insulator into a
trivial band-insulator (and Mott insulators and other phases of spin models) without
closing the band gap by using this enlarged Hilbert space. As fermions
even in an insulator can perform virtual hops from one site to the
next, one has to include in the wave function half-integer instead of
only integer representations of SU(2). This allows to unwind, for
example, the spin-1 Haldane phase\cite{anfuso}. A general discussion
of how symmetries and their representation lead to a topological
classification of spin models can be found in
Refs.~[\onlinecite{wen11,cirac11}] and [\onlinecite{pollmann1}].

We will start by discussing the concept of symmetry and adiabatic continuity in a general setup where a low-energy Hilbert space is embedded into some larger Hilbert space. Then, we will show explicitly how a trivial initial state can be deformed into a state with 4 Majorana edge modes. Finally, we will put this result in the general concept of classifying interacting systems in term of projective symmetries and the entanglement spectrum.

\section{Adiabatic continuity and classification of groundstates}\label{classify}

The principle of adiabatic continuity can be used to classify states of matter. It can, for example, be formulated in the following way.  Consider a given symmetry group $\mathcal G_s$ and two  ground state wave functions $|\Psi_0 \rangle$ and $|\Psi_1 \rangle$ of two Hamiltonians $H^s_0$ and $H^s_1$ with this symmetry which are both defined on the Hilbert space $\HH_s$. For simplicity we consider only systems in the thermodynamic limit with a unique groundstate for periodic boundary conditions, a finite gap in the spectrum (generalizations to degenerate groundstates and even gapless systems are possible but not considered here) and short ranged interactions. Then the two wave functions belong to the same topological class if a family of Hamiltonians $\HH^s_\lambda$ exists which depend smoothly on a parameter $\lambda \in [0,1]$ in such a way that
$H^s_{\lambda=0}=H^s_0$ and $H^s_{\lambda=1}=H^s_1$. Along the path all symmetries in $\mathcal G_s$ have to be preserved and the gap of $ H^s_\lambda$ has to remain finite.

The definition given above is, however, rather restrictive when one takes into account that low-energy Hilbert spaces are only an approximation to reality. Therefore it seems useful to generalize the definition of adiabatic continuity given above by taking into account a larger Hilbert space $\HH$ in which $\HH_s$ is embedded, we consider $\HH=\HH_s \otimes \HH_e$. For example the spin-1 Heisenberg model discussed above is originally
formulated on a Hilbert space build from $S=1$ spins. But in a real material, the spin $1$ is formed by the Hund's rule coupling of the spin of two localized electrons. In this case it seems reasonable to consider for the topological classification  the larger Hilbert space $\HH$ of the fermions rather than the restricted spin-1 Hilbert space $\HH_s$.

Therefore we use the following generalization of the concept of adiabatic continuity. Two ground states.  $|\Psi_0 \rangle$ and  $|\Psi_1 \rangle$, of the Hamiltonians $H_0$ and $H_1$ defined on $\HH_s$ with symmetries $\mathcal G_s$  are in the same topologically class in $\HH=\HH_s \otimes \HH_e$ if there exist a family of gapped Hamiltionians $H_\lambda$ which smoothly interpolate between
$H_0$ and $H_1$
\begin{eqnarray}
H_{\lambda=0}&=&H_0 \otimes \id_e+\id_s \otimes H_{e,0}, \\
H_{\lambda=1}&=&H_1 \otimes \id_e+\id_s \otimes H_{e,0} \nonumber
\end{eqnarray}
where $H_{e,0}$ is a gapped Hamiltonian in $\HH_e$ (typically a simple insulator).
$H_\lambda$ is defined on $\HH$ and as above we require smoothness in $\lambda$,  short-ranged interactions, a unique ground state for periodic boundary conditions and, most importantly, all symmetries should be obeyed for arbitrary $\lambda$. 

This last requirement, that all symmetries have to be obeyed, requires some care:
we want to allow (as in the case of the spin-1 chain embedded in a fermionic Hilbert space discussed above) that the symmetry group $\mathcal G$ of $\HH$ may contain some elements (e.g. $\T^2$) which act only trivially ($\T^2=1$) on $\HH_s$. We require that the symmetries of $\mathcal G$ are always obeyed 
but $\mathcal G_s$ nevertheless differs from $\mathcal G$ as the Hilbert space
$\HH_s$ uses, e.g., only spin-1 representations with $\T^2=1$.

At the beginning and end of the adiabatic path, $\lambda=0,1$, the two Hilbert spaces $\HH_s$ and $\HH_e$ are not connected. Therefore, the unique ground state wave function at $\lambda=0$ ($\lambda=1$) is a product state of the ground state of $H_0$ and $H_{e,0}$ (of $H_1$ and $H_{e,0}$). Note that the initial Hamiltonian $H_{e,0}$ and the corresponding wave function $| \Phi_0 \rangle$, is identical to the final Hamiltonian and wave function. Therefore the adiabatic transformation takes the form
\begin{eqnarray}\label{smooth}
|\Psi_0 \rangle | \Phi_0\rangle  \xrightarrow{\rm adiabatically}  |\Psi_1 \rangle |\Phi_0\rangle.
\end{eqnarray}
If such a transformation exist, $|\Psi_0 \rangle$ and $ |\Psi_1 \rangle$ are in the same topological sector for a given embedding in $\HH$ with the symmetry group $\mathcal G$.

\section{Topological phases of spinless fermions}

As a warm-up we repeat the arguments leading to the classification of spinless superconducting fermions. We consider $N_f$ chains of spinless but superconducting fermions described  in a famous paper of Kitaev \cite{kitaev01} with the Hamiltionian 
$H^{s0}= \sum (- \frac{t_f}{2} ( f^\dagger_{j,\alpha} f^{\hphantom{\dagger}}_{j+1,\alpha}+ f^\dagger_{j,\alpha} f^\dagger_{j+1,\alpha}+H.c.) + e_f f^\dagger_{i,\alpha} f^{\hphantom{\dagger}}_{i,\alpha})$ which can be written in the form
\begin{eqnarray}\label{h0}
H^{s0}=\frac{i }{2} \sum_{ j, \alpha=1,..,N_f} \left( t_f \, b_{j, \alpha} a_{j+1, \alpha}+e_f \, a_{j, \alpha} b_{j, \alpha}\right)
\end{eqnarray}
by introducing Majorana fermions $a_{j\alpha}$, $b_{j\alpha}$ with $a_{j\alpha}^\dagger=a_{j\alpha}$,  $b_{j\alpha}^\dagger=b_{j\alpha}$ and $\{a_{j\alpha},a_{j\alpha}\}=\{b_{j\alpha},b_{j\alpha}\}=2 \delta_{ij}$ using
\begin{eqnarray}
a_{j\alpha}=f_{j\alpha}^\dagger+f_{j\alpha}, \qquad b_{j\alpha}=-i (f_{j\alpha}-f_{j\alpha}^\dagger),
\end{eqnarray}
Under time reversal one therefore obtains $a_{j\alpha} \to a_{j\alpha}$ while 
$b_{j\alpha} \to -b_{j\alpha}$.
For $t_f=0, e_f>0$, Majorana fermions on the same site are paired forming a trivial band insulator. In contrast, for   $e_f=0, t_f>0$ Majorana fermion on the links are bound to each other. For an open chain with $N$ sites, $j=1,...,N$, this implies that both on the left side and the right side of the chain  $N_f$ Majorana fermions, $a_{1,\alpha}$ and $b_{N,\alpha}$, respectively, are not bound and form in total $2^{N_f}$ zero energy states (in the absence of interactions) separated by a gap from all other excitations. These zero energy states remain stable for finite $e_f< t_f$. 
Their stability in the absence of interactions is related to the fact that it is not possible to write down a quadratic term in $a_{1\alpha}$ which is both hermitian and time-reversal invariant as terms like $i a_{1 \alpha} a_{1 \alpha'}$ are odd under $\T$.  Therefore one can classify non-interacting systems by the number of Majorana edge modes, i.e. by $\Z$.

For an interacting systems, it is instructive to investigate for $N_f=4$ the consequence of the interaction term 
\begin{eqnarray}
H^{s,\rm int}=- g \sum_{i} (a_{i1} a_{i2} a_{i3} a_{i4} +b_{i1} b_{i2} b_{i3} b_{i4} + H.c. )
\end{eqnarray}
This Hamiltonian splits the $2^2=4$ states spanned by the four
Majorana states $a_{i1}, ... , a_{i4}$ ($b_{i1}, ... , b_{i4}$)  into
two doublets. Each of the two doublets has the properties of a
spin-1/2. This implies that one can define for the doublet with the
lower energy three pseudo-spin operators $\tilde S^{n}_{ia}$ ($\tilde S^{n}_{ib}$), $n=x,y,z$ using
\begin{eqnarray}\label{pseudo}
\tilde S^n_{ia}=\frac 1 2 \tilde a^\dagger_{i \sigma} \sigma^n_{\sigma \sigma'} \tilde a_{i \sigma'}, \qquad \tilde S^n_{ib}=\frac 1 2 \tilde b^\dagger_{i \sigma} \sigma^n_{\sigma \sigma'} \tilde b_{i \sigma'}
\end{eqnarray}
with $\tilde a^\dagger_{i \uparrow}=\frac 1 2 (a_{i 3} + i a_{i 4})$, $\tilde a^\dagger_{i \downarrow}=\frac 1 2 (a_{i 1} + i a_{i 2})$  and $\tilde b^\dagger_{i \uparrow}=\frac 1 2 (b_{i 3} + i b_{i 4})$, $\tilde b^\dagger_{i \downarrow}=\frac 1 2 (b_{i 1} + i b_{i 2})$, respectively. Here one has to note that the operators $\tilde a_{i \sigma}$ and $\tilde b_{i\sigma}$ can {\em not} be identified with spinful fermions as they have the wrong transformation properties under time reversal, $\T$. Nevertheless, the pseudospin  operator is a perfectly valid physical spin operator with the usual transformation properties under $\T$,  $\tilde S^n_{ia} \xrightarrow{\T}-\tilde S^n_{ia}$, and the conventional commutation relation, for example,  $[\tilde S^x_{i a},\tilde S^y_{j a}]=i  \tilde S^z_{i a} \delta_{ij}$. This simple observation immediately suggests that a state with 8 Majorana modes is not topologically stable as the coupling of two pseudospins to a pseudospin-singlet allows for a unique ground state at the edge with no remaining degeneracies. If one assumes that the edge reflects the topological classification of the bulk, this observation 'explains'  the $\Z_8$ classification obtained in Refs.~[\onlinecite{fidkowski,fidkowski2,turner11}].

\section{Unwinding four Majorana chains}
We will now couple $N_f=4$ chains of spinless fermions to spinful fermions to show that in this case one can unwind not only $8$ but also $4$ Majorana edge modes. The basic idea behind the following construction is that the coupling of a physical spin-1/2 to a pseudo spin-1/2 will also remove the edge mode. Furthermore, a real spin 1/2 is {\em not} topologically protected as it can be removed by charge fluctuations. The combination of these two facts allows to construct explicitly a Hamiltonian where the bulk wave function interpolates smoothly between an ordinary insulator and a state with 4 Majorana edge modes
of spinless fermions as in Eq.~(\ref{smooth}), see Fig.~\ref{figtrans}. 

We consider the following family of Hamiltonians parametrized by $t_f, e_f, g, t_{c1}, t_{c2},  U, \Delta$ and $J$
\begin{eqnarray}\label{hall}
H&=&H^{s0}+H^{s,\rm int}+H^e +H^{es}\\
H^e&=& - \sum_{i,\sigma=\uparrow\downarrow} ( t_{c1}  c^\dagger_{ib\sigma} c^{\hphantom{\dagger}}_{i+1a\sigma}+ t_{c2} c^\dagger_{i a \sigma} c^{\hphantom{\dagger}}_{ib\sigma} + H.c.) \nonumber\\
&&+ \sum_{i,\sigma=\uparrow\downarrow} \Delta (n_{i a}-n_{ib})
+U \sum_{i,\beta=a,b} n_{i\beta\uparrow}n_{i\beta\downarrow} \nonumber\\
H^{es}&=& J \sum_{i,\beta=a,b} \tilde{ \bf S}_{i \beta} \cdot \bf S_{i \beta} \label{hes}
\end{eqnarray}
Here $H_e$ describes a dimerized Hubbard model with $a$ and $b$ sites, a staggered potential $\Delta$ and a different hopping rate within the unit cell ($t_{c1}$) and from one unit cell to the next ($t_{c2}$). $H^{es}$ couples the spin on a given site, $S^n_{i\beta}=\frac 1 2  c^\dagger_{i \beta \sigma} \sigma^n_{\sigma \sigma'}  c_{i \beta \sigma'}$ to the pseudo-spins of Eq.~(\ref{pseudo}). The relevant symmetries of $H$ are discussed in Sec.~\ref{symmetries} below.

We consider the following path in the parameter space of $H$ parametrized by $0\le \lambda \le 2$. Initially, for $\lambda=0$, the two subsystems are decoupled and both $|\Psi_0\rangle$ and $|\Phi_0\rangle$ describe trivial band insulators of the spinless and spinful fermions, respectively, with  $\Delta,e_f=1$ and all other parameters set to $0$. In a first step, $\Delta,e_f$ are slowly switched off and the interaction parameters $g, U$ 
and, simultaneously, $J$ are switched on.
\begin{align}
{\rm for\ }  0\le \lambda \le 1:& \quad  t_f, t_{c1}=0, \quad e_f,\Delta=1-\lambda \\ & \quad g,U,J=\lambda, \quad  t_{c2}= \lambda (1-\lambda) \nonumber
\end{align}
 The hopping $t_{c2}$ is always finite for $ 0< \lambda <1$ but vanishes at the end of step 1, $\lambda=1$. During this first step, different sites $i$ always remain disconnected as $t_f, t_{c1}=0$ and no entanglement builds up (see below).
At the end of the first step, only $g, U$ and $J$ are finite and the
$a$-Majoranas are coupled only to the $a$ sites of the spinful
fermions and the $b$-Majoranas to the $b$ sites, see middle of Fig.~\ref{figtrans}.

\begin{figure}[t]
\begin{center}
\includegraphics[width=0.85 \linewidth]{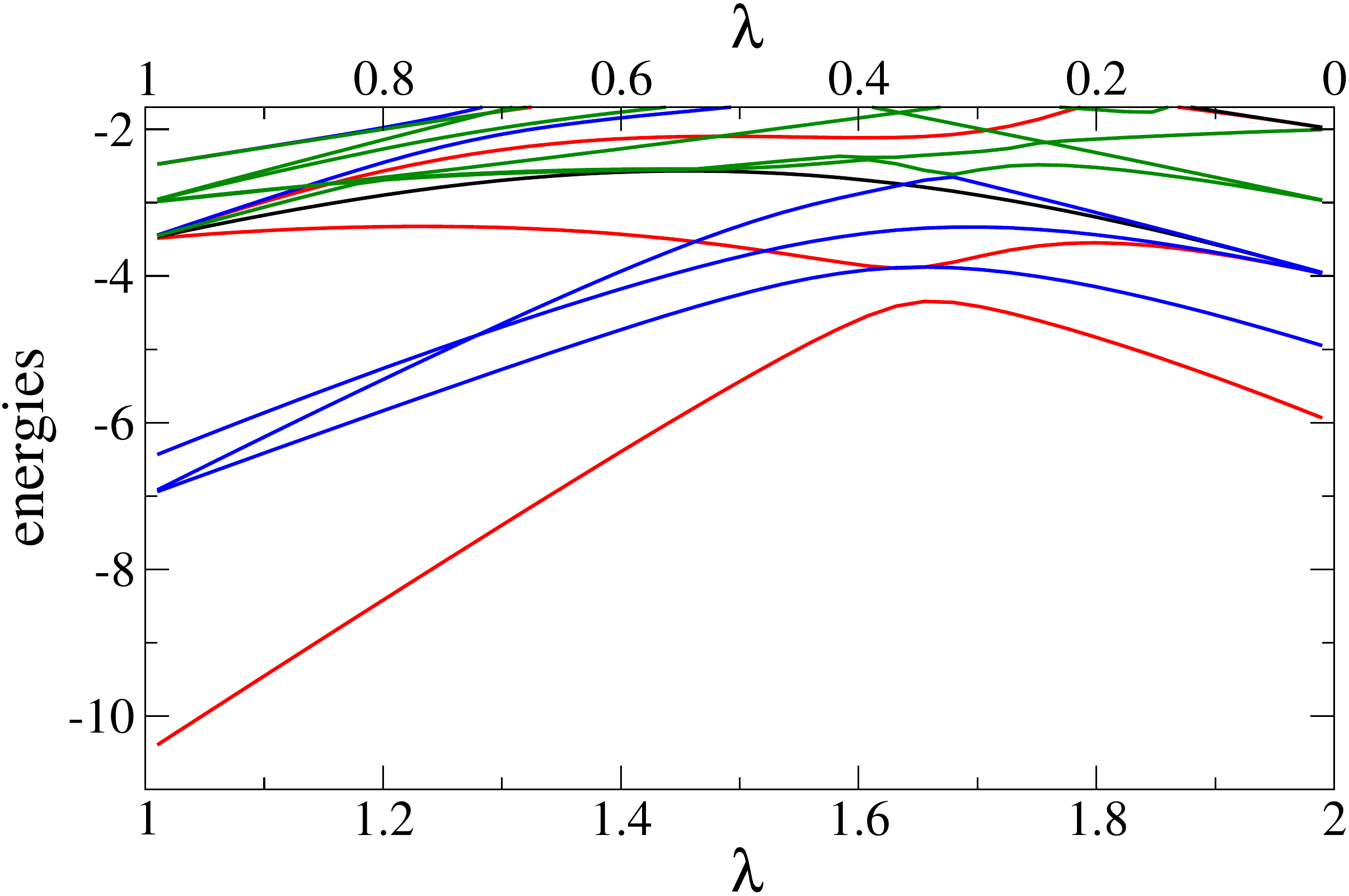}
\end{center}
\caption{(color online) Lower part of the energy spectrum of $H$ as a function of $\lambda$ for $1\le \lambda \le 2$ (lower axes labels). The spectra for $0\le \lambda \le 1$ are identical (see upper axes labels) as $E(\lambda)=E(2-\lambda)$. The ground state is unique and the gap is finite. \label{figenergies}}
\end{figure}

In the second step, the disconnected $a$ and $b$ Majoranas are connected again in a protocol which just reverses the first step up to one decisive difference: this time, $a$ and $b$ Majoranas on {\em neighboring} sites and not on the same site $i$ are connected as can be seen in Fig.~\ref{figtrans}.

\begin{align}
{\rm for\ }  1\le \lambda \le 2:& \quad  e_f, t_{c2}=0, \quad t_f,\Delta=\lambda-1 \\ & \quad g,U,J=2-\lambda, \quad  t_{c1}=  (1-\lambda) (2-\lambda)\nonumber
\end{align}
Both during the first and the second step, the Hamiltonian splits into blocks of just 2 spinful and 4 spinless fermionic sites and it is easy to diagonalize the $2^8 \times 2^8$ dimensional Hamiltonian numerically to show that the bulk gap always remains finite, see Fig.~\ref{figenergies}.

At the end of this procedure, $\lambda=2$, all interactions are switched off, spinful and spinless fermions are disconnected. While the wave function of the spinful fermions has back its original form $|\Phi_0\rangle$ describing a band insulator, the wave function of the spinless fermions is now characterized by four zero-energy Majorana states at the edge. Therefore, we have been able to adiabatically connect two wave functions which in the absence of the coupling to a larger Hilbert space would live in different topological sectors.

\section{Entanglement spectrum and topological stability}

Instead of investigating the edge modes of a finite system, a very
useful (and by now rather fashionable) approach is to investigate the
bipartite entanglement of a subsystem and, especially, the
entanglement spectrum
\cite{haldane,pollmann1,thomale,turner11,laeuchli}. The entanglement
spectrum is obtained by studying the reduced density matrix of a
subsystem, $\rho_s={\rm tr}_E |\psi\rangle \langle \psi |$, when the
rest of the system, the environment is traced out. Writing
$\rho_s=e^{-\tilde H}$, one defines the entanglement spectrum using
the eigenvalues $E_n$ of $\tilde H$. Equivalently, one can consider
the Schmidt decomposition of the groundstate wave function $
|\psi\rangle$ into states of the system and environment of the form
$|\psi\rangle=\sum_n e^{-E_n/2} |n,S\rangle|n,E\rangle$ where
$|n,S\rangle$ and $|n,E\rangle$ are states in the system and the
environment, respectively.

\subsection{Level crossings in the entanglement spectrum}

\begin{figure}[t]
\begin{center}
\includegraphics[width=0.95 \linewidth,clip]{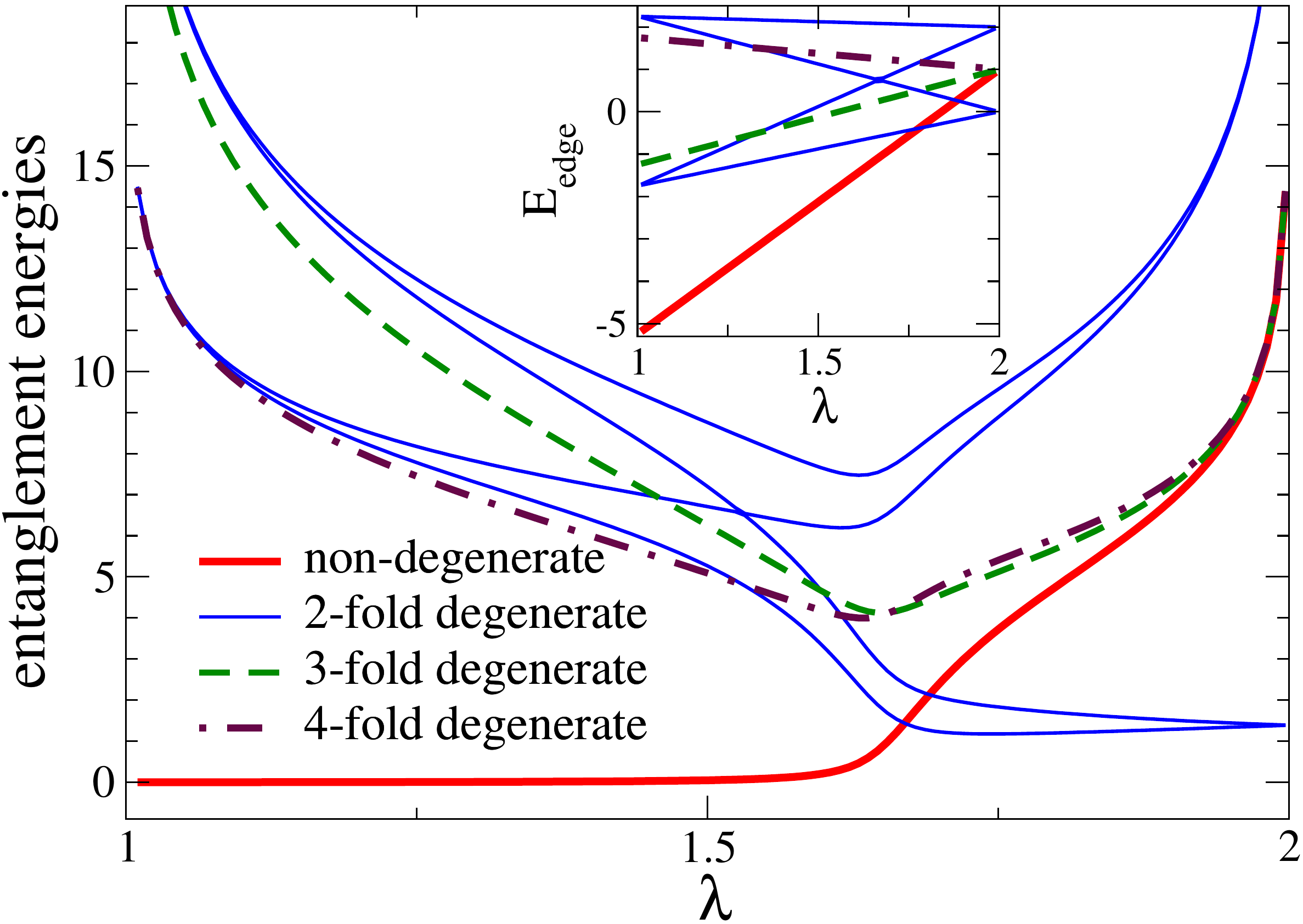}
\end{center}
\caption{(color online) Entanglement spectrum as a function of $\lambda$ for $1\le \lambda \le 2$ for an infinite system cut into two parts. It evolves from an unentangled state for $\lambda \le 1$ ($E=\ln 1=0$) towards a four-fold degenerate level ($E=\ln 4$) at $\lambda=2$ reflecting four Majorana edge states. The qualitatively different entanglement spectra are connected by a simple level crossing. Inset: The edge spectrum of an open chain also shows a simple level crossing of a unique ground state and a doubly degenerate state which merges with two other states to form a zero-energy four-fold degenerate state at $\lambda=2$.  \label{figEntanglement}}
\end{figure}

To calculate the entanglement spectrum, we consider an infinite system cut into two parts (dashed line in Fig.~\ref{figtrans}). For $\lambda\le 1$ there is no entanglement as the Hamiltonian has no matrix elements connecting the left and right side of the cut. Therefore there is only one non-vanishing eigenvalue $1$ of $\rho_s$ and the corresponding entanglement energy is zero, $E=\ln 1=0$.
For $\lambda=2$, in contrast, the cut affects four Majorana bonds (see Fig.~\ref{figtrans}) and the entanglement spectrum therefore consists of $2^{4/2}=4$ degenerate states with entanglement energy $E=\ln 4$.
Fig.~\ref{figEntanglement} shows how the system interpolates between the two spectra. Starting from the 4-fold degenerate, non-interacting state at $\lambda=2$,
arbitrarily small interactions splits the four-fold degenerate state into two doublets
when $\lambda$ is reduced. The degeneracy of these doublets is symmetry protected (see below) but when $\lambda$ is further reduced there is a simple level crossing with a non-degenerate state which obtains more and more probability in the reduced density matrix (i.e., it moves closer to $\ln 1$=0) until it becomes the only
remaining state at $\lambda=1$.  

While the entanglement spectrum characterizing probabilities within the reduced density matrix and  the spectrum of edge modes are physically distinct quantities, they share nevertheless many similarities \cite{haldane,pollmann1,thomale,turner11}. This is also the case here:  the inset of Fig.~\ref{figEntanglement} shows the change of the edge spectrum as a function of $\lambda$. While quantitatively the entanglement and the edge spectrum are completely different, qualitative features, i.e., the characteristic level degeneracies and band crossings, can be identified.

\subsection{Global symmetries} \label{symmetries}
The classification of symmetry protected topological states relies on the 
precise symmetries of the system. In this section we will therefore discuss the relevant symmetries both of the full- and the low-energy Hilbert space.

The set of symmetry operations relevant for our discussion is 
\begin{eqnarray}
\label{groupG}
\mathcal G=\{\id, \T, \Q_c, \T \Q_c, \Q, \T \Q, \Q_c \Q, \T \Q_c \Q  \}
\end{eqnarray} 
where $\Q=(-1)^{\sum_{i \alpha} f^\dagger_{i \alpha} f^{\phantom{\dagger}}_{i \alpha}}$ and $\Q_c=(-1)^{\sum_{i \sigma} c^\dagger_{i \sigma} c^{\phantom{\dagger}}_{i \sigma}}$ are the parity operators for the 
spinless and spinful fermions, respectively, with $\Q_c^2=\Q^2=\id$ and all operators commute with each other (further symmetries, e.g. inversion, change the classification \cite{pollmann1}).

While for the time-reversal invariant spinless fermions one has $f_{i\alpha}^\dagger \xrightarrow{\T} f_{i\alpha}^\dagger$, the spinor index $\sigma$ of spinful fermions transform with the Pauli matrix $\sigma^y$,
$c_{i\beta \uparrow}^\dagger \xrightarrow{\T} i c_{i\beta \downarrow}^\dagger$
and $c_{i\beta \downarrow}^\dagger \xrightarrow{\T}- i c_{i\beta \uparrow}^\dagger$. As a consequence the operator $\T^2$ is  $\id$ when acting on $\HH_s$ but $-\id$ for states in $\HH_e$ with an odd number of c-fermions (as $\T i=-i \T$).  Note also that ${\bf S}_{i\beta} \xrightarrow{\T} -{\mathbf S}_{i\beta}$ and $\tilde {\bf S}_{i\beta} \xrightarrow{\T} -\tilde{\mathbf S}_{i\beta}$ under time reversal.

Therefore one obtains $\T^2=\Q_c$ and $\mathcal G$ can be identified with $\Z_4 \times \Z_2$. 

The low-energy Hilbert space $\HH_s$ defined by the spinless fermions, however,
is build up from representations of the group $\mathcal G$
 where $Q_c=\id$ and the remaining non-trivial group elements in this subspace,
\begin{eqnarray}
\mathcal G_s=\{\id,\T,\Q,T \Q\},\label{groupGs}
\end{eqnarray}
can be identified with $\Z_2 \times \Z_2$ as $\T^2=\id$ in this subspace. While considering only 
$\mathcal G_s$, one obtains the $\Z_8$ classification of interacting 
Majorana chains as proven in Ref.~[\onlinecite{fidkowski,fidkowski2,turner11}], 
we have shown above that four edge Majorana states are not stable when considering the enlarged Hilbert space (and therefore the group $\mathcal G$).

\subsection{Projective symmetries and topological classification}

We will now show that the absence of topological projection for $n=4$
Majorana edges (and the presence of such a protection for $n=1,2,3$)
follows in a straightforward way from existing classification schemes
of interacting one-dimensional systems\cite{turner11,fidkowski2}. In
the following we will closely follow the arguments of Turner, Pollmann
and Berg\cite{turner11} to investigate why their $\Z_8$ classification
scheme has to be modified in the presence of spinful fermions and why
states with $0, 1, 2$ and $3$ Majoranas still remain topologically
stable. To be precise, we will {\em not} try to
give a complete classification of spinful and spinless electrons in
one dimension. Within the spirit of the discussion in section
\ref{classify}, we ask instead the question how interacting
time-reversal invariant spinless fermions are classified in the
presence of interactions if they are coupled to a {\em trivial}
``vacuum'' made out of band-insulating spinful fermions. We will only
sketch the argumentation of Ref.~[\onlinecite{turner11}] and refer to
this excellent paper for further details.

The basic idea for the classification\cite{turner11} is to investigate how symmetry operations affect the entanglement spectrum (or more precisely, the eigenstates of the reduced density matrix). We define our ``system'' as a finite but very large piece cut out of an infinite system. In contrast to the previous section, the entanglement therefore arises now from two cuts, a left (L) and a right (R) cut.  Turner, Pollmann and Berg\cite{turner11}  realized two important facts: (i)~Each symmetry operation $S_i$ acting on $\rho_{red}$ splits naturally into two independent parts, $S_i \propto S_{iL} S_{iR}$, affecting only properties close to the left and right part. (ii)~Classifying the possible algebras of the ``splitted'' symmetry operations $S_{iL}$ and $S_{iR}$  allows to classify the interacting system.
To split $\T$ into two parts, one has to take into account that it is an antiunitary operator. This can, for example, be done by defining the complex conjugation operator $K$ relative to some given basis. Then $\T$ can be written\cite{turner11} as $\T=U_L U_R K$ with unitary operators $U_{L/R}$ acting only close to the left or right side. This also allows to define $\T_L=U_L K$ and $\T_R=U_R K$. Defining also left and right parity operators, $\Q \propto \Q_L \Q_R$,
one can now study the algebra of $\T_{L/R}$ and $Q_{L/R}$.

Consider first the case {\em without} spinful fermions, i.e. the
situation studied in
Refs.~[\onlinecite{fidkowski,fidkowski2,turner11}] where $\mathcal G_s$ defines the relevant symmetry transformations. In this case the
global symmetries obey the three equations $\Q^2=\id$, $\T^2=\id$ and
$\Q \T=\T \Q$.  This allows for $2^3=8$ different
options\cite{turner11}: $\Q_{L/R}$ can either change the fermion
number from even to odd or not, which implies that it can be either
bosonic ($\Q_L \Q_R=\Q_R \Q_L$) or fermionic ($\Q_L \Q_R=-\Q_R
\Q_L$). Also $\T_{L/R}$ can either be bosonic or fermionic and
$\T_L^2=\T_R^2$ can either be $\id$ or $-\id$.

If we now enlarge our Hilbert space to include spinful fermions, then $\T^2=Q_c$ is $\pm \id$ on different sectors of the full Hilbert space. For non-interacting systems this does not affect the topological classification as the different irreducible representations cannot couple to each other without breaking time-reversal symmetry: a spinful and a spinless fermion cannot hybridize. But in the presence of interaction a coupling is possible, see Eq.~(\ref{hes}). As a consequence, the state
with four Majoranas is not topologically protected any more. As shown below, that can directly be traced back to the fact that $\T^2$ (and therefore also $\T_L^2$)
has always both eigenvalues $\pm 1$. States with $0$, $1$, $2$ and $3$ Majoranas 
remain, however, stable. They belong to different symmetry classes
distinguished by whether $\Q_{L/R}$ and $\T_{L/R}$ are fermionic or bosonic. 
 
The identification of the four phases can easily be
worked out by studying how the symmetry operators act on chains with $n=0,1,2,3$ Majorana edge modes (consistent with the classification table in Ref.~[\onlinecite{turner11}]). This can be done by tracking how $Q$ and $\T$
affect the reduced density matrix in the limit of uncoupled, non-interacting Majorana chains with $t_f=1$, $e_f=0$.
Consider, for example, the case $n=2$ and a situation where $t_f=1, e_f=0$.
In this case only the four Majorana operators $a_{L1},a_{L2},b_{R1},b_{R2}$ are affected by the cut of the system ($L/R$ refers to the position at the left and the right edge of the ``system''). Defining the fermions $a_L^\dagger=\frac{1}{2} (a_{L1}+ i a_{L2})$  and $b_R^\dagger=\frac{1}{2} (b_{R1}+ i b_{R2})$ allows to build the 4-dimensional basis $|00\rangle,|01\rangle,|10\rangle,|11\rangle$  describing the occupation of the two fermions
with $a_L |00\rangle=b_R |00\rangle=0$. In this space the reduced
density matrix takes the form $\rho_{\rm red}=\frac 1 4 \id$. Parity
is given by $\Q=\Q_L \Q_R$ with bosonic operators
$Q_{L/R}=(-1)^{n_{L/R}}$ and $n_L=
a_L^\dagger a^{\hphantom{\dagger}}_L$,  $n_R=
b_R^\dagger b^{\hphantom{\dagger}}_R$. Computing
$\T \Q_L \T^{-1}=(-1)^{a^{\hphantom{\dagger}}_L a_L^\dagger}=- \Q_L$
shows that $\T Q_L=-Q_L \T$. Therefore $\T_L$ has to be fermionic. The
explicit calculation shows that in the chosen basis $\T=a_{L1} b_{R2}
K$ with a fermionic $\T_L=a_{L1} K$ where $K$ is defined by $K
\sum_{i,j=0,1} \alpha_{ij} |ij\rangle = \sum_{i,j=0,1} \alpha_{ij}^* |ij\rangle$. 

One can repeat this argument for arbitrary $n$:
$\Q_L$ is the product of $n$ edge Majoranas and therefore fermionic for odd $n$ and bosonic for even. Similarly, $\T_L$ is fermionic for $n=1,2$ but bosonic for $n=0,3$. A state with fermionic $\T_L$ or $\Q_L$ cannot  be transformed smoothly into a state with bosonic $\T_L$ or $\Q_L$ and therefore the states with
$n=0,1,2,3$ belong to four different topological sectors.

For $n=4$, where both $\Q_{L/R}$ and $\T_{L/R}$ are bosonic, we have shown that an adiabatic path to $n=0$ does exist which uses a simple level crossing of
a non-degenerate groundstate ($\T_L^2=1$) and a Kramer's doublet ($\T_L^2=-1$), see Fig.~\ref{figEntanglement}. Such a level crossing is not possible if one has globally $\T^2=1$. As soon as one allows that  $\T_L^2$ has simultaneously the eigenvalues $+1$ and $-1$, it is not possible to fulfill the condition that $\T^2=1$ for {\em all} its eigenstates. 

We have therefore shown, that states with $0$, $1$, $2$ and $3$ 
Majorana modes at a given edge are topological stable in our system while one can deform
a state with $n$ Majorana modes at the edge in a state with $n \mod 4$ Majorana modes. Note, however, that we have only focused on the low-energy Hilbert space of our subsystem while we have not investigated the possibility of other topological
states involving the spinful fermions.

\section{Conclusions}

With this paper we have added one more example to the (short) list
of cases where interactions enforce to modify the classification of non-interacting topological 
insulators and superconductors. Our simple examples can be used to demonstrate several points: (i) The topological 
classification of a low-energy Hilbert space can be insufficient in the sense that
in a larger Hilbert space it is possible to unwind topological states of the low-energy space. This can happen if the larger Hilbert has different symmetries in the following sense: the irreducible representations forming the low-energy Hilbert space lead to a different symmetry group. In our specific example, the key was that one of the symmetries of $\HH$  ($\T^2$) had a trivial representation ($\T^2$=1) in $\HH_s$ and the groups $\mathcal G$ in Eq. (\ref{groupG}) and $\mathcal G_s$ in Eq. (\ref{groupGs}) differ.
Another example for this are spin models embedded in fermionic Hilbert spaces \cite{anfuso}. (ii) When one uses entanglement to characterize a system, it is
not sufficient to concentrate on the lowest states in the spectrum. 
Levels with other degeneracies and symmetry properties may cross
with the low energy levels and completely rebuild the entanglement properties
without closing any bulk gap. To exclude this, one has to classify {\em all} those states
in the entanglement spectrum which survive in the thermodynamic limit (i.e. for the reduced density matrix of an infinitely large subsystem in an infinitely large total system). (iii) Finally, our example demonstrates  the validity power of the existing classification schemes
of interacting one-dimensional system \cite{turner11,fidkowski2}.

An important ingredient of our analysis was the presence of spinless fermions. 
Building a system well described by spinless fermions interacting with spinful fermions
is not simple but at least in principle possible. Spinless fermions can, for example,
be realized with ultracold alkaline atoms with a polarized nuclear spin. While polarizing
the nuclear spin nominally breaks $\T$-invariance, due to the closed
electronic shell of alkaline atoms
this has practically no consequences for the interaction with other particles.

An important challenge for the future is to find a complete classification scheme for interacting spinful electrons in higher dimensions. It will also be useful to search for higher-dimensional model systems which can be used to study in concrete examples how interactions modify the topological classification, see for example Ref.~[\onlinecite{higher}].

\acknowledgements
I acknowledge useful discussions with  E. Altman, M. Garst, E. Sela, S. Trebst,
M. Zirnbauer and, expecially, F. Pollmann and T. Quella. I thank
the DFG (SFB 608 and SFB/TR 12) for financial support.


\begin{thebibliography}{99}


\bibitem{Kane2005}{%
C.\ L.\ Kane and E.\ J.\ Mele, Phys.\ Rev.\ Lett.\ \textbf{95}, 226801 (2005).
}

\bibitem{Bernevig2006}{%
B.\ A.\ Bernevig, T.\ L.\ Hughes, S.\ C.\ Zhang, Science \textbf{341}, 1757 (2006).
}


\bibitem{Fu2007}{%
L.\ Fu, C.\ L.\ Kane, and E.\ J.\ Mele, Phys.\ Rev.\ Lett.\ \textbf{98}, 106803 (2007).
}

\bibitem{Moore2007}{%
J.\ E.\ Moore and L.\ Balents, Phys.\ Rev.\ B \textbf{75}, 121306 (2007).
}

\bibitem{Roy2009}{%
R.\ Roy, Phys.\ Rev.\ B \textbf{79}, 195322 (2009).
}

\bibitem{rev1} M. Z. Hasan, and C. L. Kane, Rev. Mod. Phys. {\bf 82}, 3045 (2010).


\bibitem{rev2} X.-L. Qi and S. -C. Zhang, Rev. Mod. Phys. {\bf 83}, 1057 (2011).


\bibitem{Koenig2007}{%
M.\ K\"onig \textit{et al.}, Science \textbf{318}, 766 (2007).
}

\bibitem{Hasan2009}{%
D.\ Hsieh \textit{et al.}, Nature \textbf{460}, 1101 (2009).
}

\bibitem{Bruene2011}{%
C.\ Br\"une \textit{et al.}, Phys.\ Rev.\ Lett.\ \textbf{106}, 126803 (2011).
}

\bibitem{schnyder08} A. P. Schnyder, S. Ryu, A. Furusaki, and A. W. W. Ludwig, Phys.
Rev. B {\bf 78}, 195125 (2008).
\bibitem{zhang08} X.-L. Qi, T. L. Hughes, and S.-C. Zhang, Phys. Rev. B {\bf 78}, 195424
(2008).
\bibitem{kitaev09} A. Yu Kitaev, AIP Conf. Proc. 1134, 22  (2009).

\bibitem{altland} A. Altland and M. R. Zirnbauer, Phys. Rev. B 55, 1142 (1997).




\bibitem{Qi2008}{%
X.-L.\ Qi, T. L. Hughes, S.\ C.\ Zhang, Phys.\ Rev.\ B \textbf{78}, 195424 (2008).
}



\bibitem{fidkowski}
L. Fidkowski and A. Kitaev, 
Phys. Rev. B {\bf 81}, 134509 (2010).

\bibitem{turner11} A. M. Turner, F. Pollmann, E. Berg,
Phys. Rev. B {\bf 83}, 075102 (2011).

\bibitem{fidkowski2} 
L. Fidkowski and A. Kitaev, Phys. Rev. B 83, 075103 (2011).


\bibitem{wen11} 
Xie Chen, Zheng-Cheng Gu, Xiao-Gang Wen, Phys. Rev. B {\bf 83}, 035107 (2011).

\bibitem{cirac11} N. Schuch, D. P\'erez-Garc\'ia, and I. Cirac, 
Phys. Rev. B {\bf 84}, 165139 (2011).

\bibitem{anfuso}
F. Anfuso and A. Rosch, Phys. Rev. B {\bf 75}, 144420 (2007);
F. Anfuso and A. Rosch, Phys. Rev. B {\bf 76} , 085124 (2007);
A. Rosch, Eur. Phys. J. B {\bf 59}, 495 (2007). 



\bibitem{pollmann1} F. Pollmann, A. M. Turner, E. Berg, and M. Oshikawa,
Phys. Rev. B {\bf 81}, 064439 (2010).

\bibitem{kitaev01} A. Kitaev, Phys. Usp. {\bf 44}, 131 (2001).




\bibitem{haldane} H. Li and F. D. M. Haldane, Phys. Rev. Lett. {\bf 101}, 010504
(2008).


\bibitem{thomale} R. Thomale, D. P. Arovas, and B. A. Bernevig, Phys. Rev. Lett. {\bf 105},
116805 (2010); R. Thomale, A. Sterdyniak, N. Regnault, and
B. A. Bernevig,
 Phys. Rev. Lett. {\bf 104}, 180502 (2010).


\bibitem{laeuchli} A. M. L\"auchli, E. J. Bergholtz, J. Suorsa, and M. Haque, Phys. Rev. Lett. {\bf 104}, 156404 (2010).

\bibitem{higher} Xiao-Liang Qi, preprint arXiv:1202.3983v2;
Zheng-Cheng Gu, Xiao-Gang Wen, preprint arXiv:1201.2648v1;
Hong Yao, Shinsei Ryu, preprint arXiv:1202.5805v1. Shinsei Ryu,
Shou-Cheng Zhang, Phys. Rev. B {\bf 85}, 245132 (2012).

\end{thebibliography}

\end{document}